# Discrete stochastic charging of aggregate grains

Lorin S. Matthews[1], Babak Shotorban[2], and Truell W. Hyde[1]

[1] *Center for Astrophysics, Space Physics, and Engineering Research One Bear Place 97310, Waco, TX 76798*
[2] *Department of Mechanical and Aerospace Engineering, The University of Alabama in Huntsville, Huntsville, AL 35899*

Abstract

Dust particles immersed in a plasma environment become charged through the collection of electrons and ions at random times, causing the dust charge to fluctuate about an equilibrium value. Small grains (with radii less than 1 μm) or grains in a tenuous plasma environment are sensitive to single additions of electrons or ions. Here we present a numerical model that allows examination of discrete stochastic charge fluctuations on the surface of aggregate grains and determines the effect of these fluctuations on the dynamics of grain aggregation. We show that the mean and standard deviation of charge on aggregate grains follows the same trends as those predicted for spheres having an equivalent radius, though aggregates exhibit larger variations from the predicted values. In some plasma environments, these charge fluctuations occur on timescales which are relevant for dynamics of aggregate growth. Coupled dynamics and charging models show that charge fluctuations tend to produce aggregates which are much more linear or filamentary than aggregates formed in an environment where the charge is stationary.

## I. INTRODUCTION

One of the fundamental processes which occurs in a complex plasma environment is the charging of dust grains, nm – mm-sized solid particulates immersed in the plasma. Dust particles acquire charge through direct collisions with electrons and ions, and in most cases will become negatively charged due to the initial large flux of electrons to the uncharged grain. In some plasma environments, secondary electron emission (SEE) or photoemission can contribute to the charging process, and dust grains can become positively charged.

In all of these cases, the charge is gained or lost in discrete units of elementary charge at random time intervals, with the charge on a grain fluctuating about the average equilibrium charge. It was predicted theoretically [1], and later confirmed through numerical simulations of the fluctuations [2], [3] and solutions to master or Fokker-Plank (FP) equations [4]–[7] that the time-averaged charge on the grain is linearly proportional to the grain radius, while the standard deviation is proportional to the square root of the average charge. Thus, the magnitude of the charge fluctuations relative to the equilibrium charge decreases as the grain size increases.

For example, under typical experimental conditions using a RF plasma, grains collect ~1,000 elementary charges per micron of radius, with a charge fluctuation on the order of 1% of the equilibrium charge [4]. However, in cases where the grain size is small, the plasma is cold or tenuous, or the dust density is large enough to remove a significant fraction of the electrons from the plasma, the average charge can be as small as hundreds or even tens of electrons. In this case, charge fluctuations can become a significant fraction of the equilibrium charge [4]. At the same time, the characteristic time scale for these charge fluctuations can become comparable to the time scales of the dynamic processes affecting the dust [3], [5].

A particular case where discrete stochastic charge (DSC) fluctuations can play a role in the dynamic response of the dust grains is the growth of the dust through aggregation. In the low pressure plasmas used for etching or chemical-vapor deposition, the formation and growth of fine particles can be detrimental to the system [8], although a similar process is often utilized to study the nucleation and growth process [9], [10]. The production of ultrafine powders with prescribed size ranges also takes advantage of thermal nucleation and particle charging, such as the nanoparticles produced for use in inhalation toxicity studies [11]. In protoplanetary disks, aggregate growth from small particles is a necessary preliminary step in the formation of the larger bodies which eventually form planetary systems. In each of the above, charge fluctuations can affect the coagulation rate, aggregate porosity, and maximum grain size [12]. In some environments, fluctuating charge on very small grains can even allow them to become positively charged [2], [3] leading to oppositely charged grains in the overall population and sometimes creating runaway growth [13], [14].

Our previous work on stochastic charging extended the stochastic charging model to non-spherical aggregate grains by treating charge as a continuous variable with charging time steps set as a fixed fraction of the equilibrium charging time [12]. As noted above, however, in many cases the grain charge is small enough that the gain or loss of charge should be quantized in units of the elementary charge, which also requires predicting the (random) elapsed time for the addition of a charged particle. In this study, a methodology is developed to model discrete fluctuations over the surface of a dust particle, with the addition of electrons and ions occurring at random times at random locations on the dust surface.

The paper is organized as follows: Section 2 provides an overview of the charging currents calculated from OML theory, and describes how these currents are used to calculate the electron and ion currents to points on a grain surface. Section 3 describes how the currents to the surface points are used to calculate the stochastic variation of the charge, by determining the random elapsed time, the charge (electron or ion) to be added and the location on the grain surface for the addition. As shown in Section 4, the model is validated by applying it to spherical grains and comparing the characteristics of the charge fluctuations to those found from previous models (which treat the grain surface as an isopotential), before applying the method to aggregate grains. Finally, a dynamic model of collisional grain growth including the effects of stochastic charge variations is presented in Section 5.

## II. CHARGING CURRENTS

In this paper, we limit our analysis to grain charging through primary electron and ion currents created by electron and ion collisions with the dust surface. This method can easily be extended to include secondary charging effects such as photoemission and secondary electron emission [15]. Inherent assumptions we make in the following are that the particle radius is smaller than the Debye length of the plasma, which is smaller than the mean free path of the plasma particles, $a \ll$

$\lambda_D \ll \lambda_{mfp}$, and that the interparticle distance between dust grains is larger than the Debye length, such that the charge on a grain is independent of other grains.

## A. OML charging currents

The charge on a dust particle is commonly determined using orbital-motion-limited theory to find the primary electron and ion currents to the grain as a function of the grain potential [16]. The current density $J_s$ to a point on the surface of a grain is determined by the flux of particles with enough energy to overcome the coulomb potential barrier to reach the surface

$$J_s = q_s n_s \int_{v_{\min}}^{\infty} v_s^3 f(v_s) \sigma(v_s) dv_s \iint \cos\gamma \, d\Omega \qquad (1)$$

where $n_s$ is the plasma density of species $s$ very far from the grain, $q_s$ the charge of the incoming plasma particle of mass $m_s$ and temperature $T_s$, $v_s$ the velocity of the incoming plasma particle with a velocity distribution $f(v_s)$, and $\sigma(v_s)$ the effective cross section of the charged target [16], which is $\sigma(v_s) = \left(1 - 2q_s\phi_d / m_s v_s^2\right)$ for particles with $2q_s\phi_d / mv_s^2 \leq 1$ and zero otherwise. The lower limit of integration for the particle velocity is the minimum velocity required for a charged plasma particle to reach a point on the surface of a dust grain having potential $\phi_d$. Thus, the minimum velocity is either zero, when the plasma species and dust have opposite charge, or $v_{\min} = \sqrt{2q_s\phi_d / m_s}$, for plasma species and dust of the same charge polarity. In the integration over the angles, $\gamma$ is the angle between the velocity vector and the surface normal, and $d\Omega = \sin\theta \, d\theta \, d\phi$ the solid angle from which the plasma particle approaches the surface. Assuming that the electrons and ions have Maxwellian velocity distributions characterized by the temperatures $T_e$ and $T_i$, respectively, Eq. 1 can be integrated easily for a point on the surface of a spherical grain. The current to the grain surface is then found by multiplying the current density by the surface area, yielding

$$\begin{aligned}
I_e &= I_{0e} \exp\left(\frac{e\phi_d}{kT_e}\right) & \phi_d &< 0 \\
I_i &= I_{0i} \left(1 + \frac{q_i\phi_d}{kT_i}\right) & \phi_d &< 0 \\
I_e &= I_{0e} \left(1 + \frac{e\phi_d}{kT_e}\right) & \phi_d &> 0 \\
I_i &= I_{0i} \exp\left(-\frac{q_i\phi_d}{kT_i}\right) & \phi_d &> 0
\end{aligned} \qquad (2)$$

The coefficients $I_{0e}$ and $I_{0i}$ represent the currents to an uncharged grain of radius $a$; assuming the plasma is isotropic and not flowing past the grain, these are

$$I_{0s} = 4\pi a^2 n_s q_s \left(\frac{kT_s}{2\pi m_s}\right)^{1/2} \tag{3}$$

Initially, the electron current to an uncharged grain is greater than the ion current. As the grain accumulates negative charge, slower moving electrons will not have the energy required to reach the grain surface, while ions continue to be attracted to the grain. Eventually an equilibrium potential is reached when the electron and ion currents are equal, where $\phi_d$ is the solution to

$$\exp\left(\frac{e\phi_d}{kT_e}\right) = \frac{n_e}{n_i}\left(\frac{m_e T_i}{m_i T_e}\right)^{1/2}\left(1 - \frac{e\phi_d}{kT_i}\right). \tag{4}$$

The equilibrium charge on a spherical grain is then given by

$$Q = 4\pi\epsilon_0 a\phi_d \tag{5}$$

### B. OML_LOS: Charging of non-spherical grains

Non-spherical grains, such as aggregates comprised of spherical monomers, have a varying surface potential. (This can be true for a spherical dielectric grain as well, if the rate of collection of electrons and ions is fast compared to the time scale for charge recombination on the surface, but not so fast as to keep all points on the surface in equilibrium.) In addition to the varying potential, the trajectories of incoming plasma particles to some points on the surface may also be blocked by other monomers in the aggregate. OML theory requires that all positive energy orbits connect back to infinity, and not originate from another point on the grain [17].

The current densities at various points on the grain surface can be found numerically [18]. To accomplish this, the surface is divided into patches surrounding points which are uniformly and randomly distributed over the surface of each sphere (see Figure 1). The potential at the center of each patch $\phi_p$ can then be calculated from the charge $q_j$ on all other surface points (at distance $r_{pj}$ away) and a patch centered about the point itself, $\phi_p = \sum_j q_j / r_{pj} + \phi_c$. The potential at the center of the patch, $\phi_c = q_p \sqrt{2/(1-\cos\theta)} / 4\pi\epsilon_0 a$, is approximated by the potential at the center of a spherical cap with surface charge density $\sigma = q_p / 2\pi a^2(1-\cos\theta)$, where $\theta$ is equal to the average angular separation between the points. As each patch on the surface of a sphere is actually a polygon, the accuracy of this approximation increases with the number of patches used in the simulation.

The current density incident on each patch is determined by numerical integration of Equation 1. The integral over the speed $v_s$ is exact, and can be calculated once the potential $\phi_p$ is known. The integral over the angles is approximated by breaking up the solid angle into many sections characterized by test directions $\hat{t}$ and determining which lines of sight (LOS) are blocked by other monomers in the aggregate, as illustrated in Figure 1. The LOS factor $LOS_p = \iint \cos\gamma \, d\Omega \approx \sum_t \cos\gamma_t \, \Delta\Omega$, is obtained by summing over the open LOS (see Matthews et al. [18] for a complete description of this treatment). The electron and ion currents to a patch are then the same as those given by Eq 2 with $I_{0s}$ replaced by the coefficient

$$I'_{0s} = A_p n_s q_s \left(\frac{kT_s}{2\pi m_s}\right)^{1/2} LOS_p / \pi \tag{6}$$

Where $A_p$ is the area of the patch.

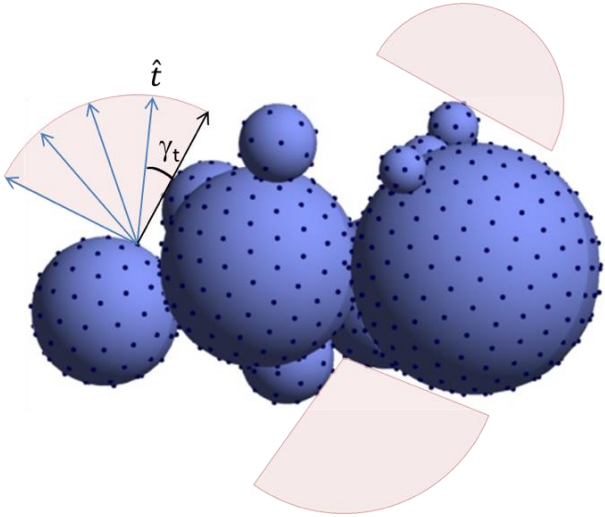

Figure 1. A 2D representation of the open lines of sight to three points on the surface of an aggregate. The open lines of sight are determined by checking many test directions $\hat{t}$ to see if they intersect other monomers in the aggregate. The angle between the surface normal vector and the test direction is $\gamma_t$.

## III. DISCRETE STOCHASTIC CHARGING METHOD

The electrons and ions in the plasma do not constitute a continuous fluid, but rather reach the grain surface at random times. The master equation for stochastic charging of aggregates developed by Matthews et al [12] treated current as a continuous variable, but here this has been modified to allow for integer increments of elementary charges collected on the surface patches.

As described in [12], a generalized form of the master equation given by Matsoukas and Russell [4], [5] and Shotorban [6] can be formulated to determine the charge collected on each patch, and hence the entire surface of the grain, utilizing the ion and electron currents to each patch. The set of elementary charges collected on the patches is defined by the vector $\mathbf{Z} = \{Z_1, Z_2, ..., Z_n\} \in \mathbb{R}^n$, where there are $n$ patches on the aggregate, (e.g., $Z_2$ is the number of elementary charges collected on the patch number 2). Assuming that $\mathbf{Z}$ undergoes a Markov process [19], the master equation is

$$\frac{dP(\mathbf{Z},t)}{dt} = \sum_{p=1}^{n} I_{i,p}(\mathbf{Z}-\mathbf{e}_p)P(\mathbf{Z}-\mathbf{e}_p,t) + I_{e,p}(\mathbf{Z}+\mathbf{e}_p)P(\mathbf{Z}+\mathbf{e}_p,t) - [I_{i,p}(\mathbf{Z}) + I_{e,p}(\mathbf{Z})]P(\mathbf{Z},t) \quad (7)$$

where $P(\mathbf{Z}, t)$ is the joint probability density function. In this equation, $I_{e,p}$ and $I_{i,p}$ are the electron and ion attachment *rates* (i.e., the electron and ion currents divided by the charge), to patch $p$, and $\mathbf{e}_p \in \mathbb{R}^n$ is the unit vector, e. g., $\mathbf{e}_3 = \{0, 0, 1, \ldots, 0\}$. It is assumed no charge is transferred from one patch to another.

In accordance with the master equation, the discrete stochastic method (DSM) is based on the following algorithm, which is a customized version of the stochastic simulation algorithm developed for chemical kinetics [20], [21], to calculate discrete charge fluctuations on patches. The system is initialized with the charges of patches set to $\mathbf{Z} = \mathbf{Z}_0$ at $t = t_0$, where $\mathbf{Z}_0$ is the initial condition. The attachment rates, $I_{i,p}(\mathbf{Z})$ and $I_{e,p}(\mathbf{Z})$, are found from the currents to each patch (Equation (6)), and then used to calculate the sum of the attachment rates to all patches,

$$\lambda(\mathbf{Z}) = \sum_{p=1}^{n} I_{i,p}(\mathbf{Z}) + I_{e,p}(\mathbf{Z}) \quad (8)$$

Then, a random number $r_1$ is generated from a uniform distribution with $0 \leq r_1 \leq 1$ and used to determine the time interval $\tau$ which elapses before the attachment of the next plasma particle

$$\tau = \frac{1}{\lambda(\mathbf{Z})} \ln\left(\frac{1}{r_1}\right) \quad (9)$$

The type of the plasma particle and the patch $p$ to which the particle is attached is determined by generating a second random number, $r_2$, and finding $k$, the smallest integer satisfying

$$\sum_{k'=1}^{k} I_{k'}(Z) > r_2 \lambda(\mathbf{Z}) \quad (10)$$

where we have defined

$$I_{k'}(\mathbf{Z}) = \begin{cases} I_{i,p}(\mathbf{Z}) \text{ if } k' = 2p-1 \\ I_{e,p}(\mathbf{Z}) \text{ if } k' = 2p \end{cases} \quad (11)$$

If $k$ is odd, then the attached particle is an ion and the patch to which the particle is attached is $p = (k+1)/2$. If $k$ is even, then the attached particle is an electron and the patch is $p = k/2$. Following this notation, for instance, $I_7(\mathbf{Z}) = I_{i,4}(\mathbf{Z})$ is the ion current to patch number 4 and $I_8(\mathbf{Z}) = I_{e,4}(\mathbf{Z})$ is the electron current to patch number 4. The time is then updated from $t$ to $t+\tau$, and the charge is updated from **Z** to **Z** + **e**$_p$ if the attached particle is an ion, or **Z** to **Z** - **e**$_p$ if the attached particle is an electron. The procedure is iterated until the desired length of time has elapsed.

## IV. CHARGING WITH THE DSM

### A. Validation of DSM: Spherical Grains

In the following section we compare the DSM for calculation of charge collected on patches on the surface of a grain to previous studies which modeled the stochastic fluctuation of charge on spherical grains [3], [5].

We apply the model using two different plasma conditions, the first being a typical low-temperature plasma discharge environment with the second using conditions which may be found in an astrophysical plasma such as that found in a protoplanetary disk (PPD). The two plasma environments will be referred to as *LAB plasma* and *PPD plasma*, respectively. Conditions for the LAB plasma assume singly ionized argon with electron and ion temperatures $T_e$ = 1 eV, $T_i$ = 500 K, and equal electron and ion number densities, $n_e = n_i = 10^{16}$ m$^{-3}$ [5]. The condition for the PPD plasma are chosen to represent a region of the disk where the dust density is large enough to deplete the electrons in the plasma [12], [18]. The ionized species is considered to be hydrogen with $T_e = T_i$ = 900 K, ion density $n_i = 5 \times 10^8$ m$^{-3}$, and electron density $n_e = 0.1 n_i$.

The master equation in the DSM assumes attachment rates vary linearly with charge. This condition is automatically satisfied for the ion current to a negatively charged grain, but the electron current to a negative grain varies exponentially with the potential (and hence the charge). This linearity criterion is satisfied by requiring a minimum grain radius for given plasma parameters, $a^* = e^2 / (4\pi\epsilon_0 k_b T_e)$. Here we apply this as a criterion for the minimum patch size. Taking the area of a patch to be $A_p = 4\pi a^2/n$, the radius of a patch $r_p$ can be estimated from $A_p = \pi r_p^2$, so that the number of patches $n = (2a/r_p)^2$. The linearity criterion is well satisfied if $r_p > 10a^*$ such that the maximum number of patches can be set by $n_{max} = (2a/10a^*)^2$.

The accuracy of the charge calculation from OML_LOS increases as the number of patches is increased for two reasons. First, the contribution to the patch potential from the charge on a patch itself is approximated assuming that the patch boundary is essentially circular. This assumption becomes less accurate as the number of patches decreases (the exception being that it is exact for $n = 2$). Second, for aggregate grains, the maximum number of patches used must be balanced against the minimum number of patches needed to resolve the LOS_factor. To accurately resolve the blocked lines of sight, the patch size on a monomer needs to be comparable to the size of the smallest monomer (in the case of a polydisperse distribution of monomers). In this case, $n = (2a_{max}/a_{min})^2$. This can lead to a contradiction where $n_{min} > n_{max}$, in which case a compromise must be reached. Using too many patches on the surface in the DSM method results in an over-estimation of the average charge on a particle (i.e., for a negatively charged grain, the number of electrons). However, using too few patches results in the charge being underestimated, as additional LOS are blocked. In general the minimum number of patches which yields accurate estimates of the equilibrium charge is $n = 10$.

Charging simulations were carried out for spheres with radii $a$ = 10, 20, 50, 100, 200, 500, and 1000 nm, and varying the number of patches on the surface with $n$ = 10, 20, 40, and 90. The charging simulations were run for 100,000 random time steps. Sample charging histories and probability distributions of the fluctuating charge are shown in Figure 2 for a 20 nm grain in a LAB plasma and a 100 nm grain in a PPD plasma.

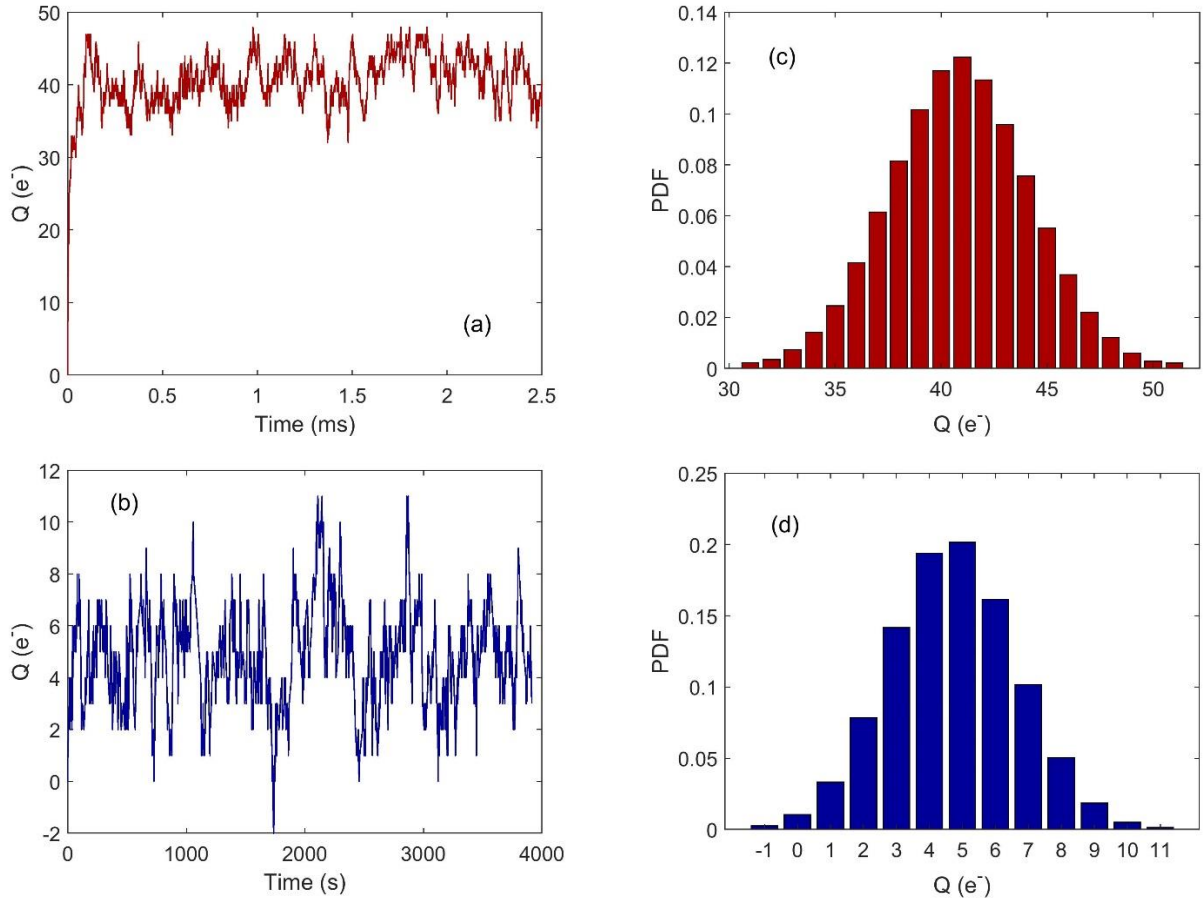

Figure 2. First 2500 time steps of the charging history (a, b) and probability distribution of the fluctuating charge (for 70,000 time steps after equilibrium is reached) (c, d). The upper panels are for a 20 nm grain with resolution $n = 20$ in LAB plasma conditions. The lower panels are for a 100 nm grain with resolution $n = 10$ in a PPD plasma.

The effect of changing the number of patches used in calculating the charge is shown in Figure 3 for several different particle radii using the two different plasma conditions, where $a^* = 1.5$ nm for the LAB plasma and $a^* = 19$ nm for the PPD plasma. The average charge on each sphere (averaged over the last 70,000 time steps) differs from the charge predicted for a sphere (Eq. 5) by less than 2%, for $r_p > 10a^*$, as shown in Figure 3. Patch sizes comparable to $a^*$ still produce reasonable results, with the average charge over-predicted by 10-15%.

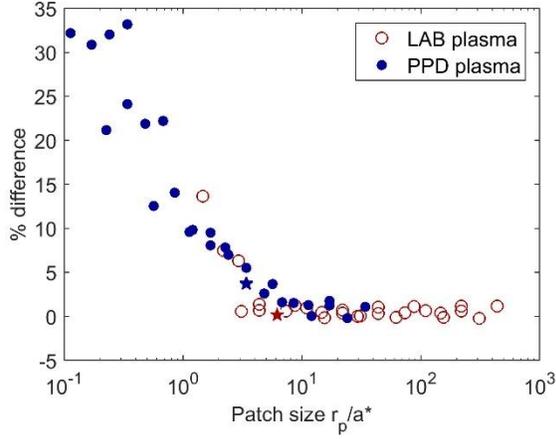

Figure 3. Percent difference between the equilibrium charge calculated with DSC and the predicted value as a function of the patch size. Points are shown for grain radii ranging from 10 nm to 1 μm using resolutions of 10, 20, 40, and 90 surface patches. The two grains shown in Figure 2 are marked with stars.

One of the main characteristics of the charge fluctuations on spherical grains is that the time scale of the fluctuations depends on the grain size, with small grains expriencing longer fluctuation times than larger grains in the same plasma environment [3], [5]. The growth and dissipation times for fluctuations of different magnitude, calculated from the time history of the charge, are denoted by symbols in Figure 4 for a 20 nm grain (LAB plasma) and a 100 nm grain (PPD plasma). The lines are analytic fits for the first-passage problem, as given by Matsoukas and Russell [5]. The characteristic fluctuation time is approximately equal to the point where the growth and dissipation curves cross. As described by Matsoukas and Russell, the fluctuation time can be calculated from

$$\tau_f = \left(\frac{4\lambda_i^2}{v_i a}\right)\frac{1}{1+T_i/T_e - \phi_d/k_B T_e} \tag{12}$$

where $\lambda_i = \left(\epsilon_0 k_B T_i / n_i e^2\right)^{1/2}$ is the ion Debye length and $v_i = \left(8 k_B T_i / \pi m_i\right)^{1/2}$ is the average speed of the ions. The excellent agreement between the analytic charging curves and those calculated using the time history generated by DSM show that the patch model is a valid technique for calculating the fluctuating grain charge.

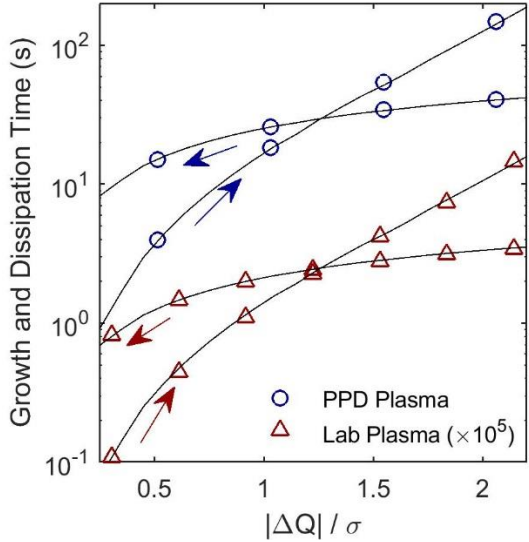

Figure 4. Characteristic times for growth (right arrows) and dissipation (left arrows) of fluctuations. The symbols are calculated from the charging time history of a 20 nm grain in a LAB plasma and a 100 nm grain in a PPD plasma. The lines are analytic results for a sphere.

## B. Application of the DSM to Aggregates

Having verified the accuracy of DSM for a sphere, this methodology is applied on aggregates of spherical grains. We test the method using three different populations of aggregates: aggregates consisting of monodisperse 500 nm and 1.0 micron spheres, as well as aggregates consisting of polydisperse spheres with monomer radii 0.5 μm ≤ a ≤ 10 μm with a size distribution typical of astrophysical environments, $n(a)da = a^{-3.5}da$. Aggregates are compared to spherical particles, by defining an equivalent radius, $R_\sigma$, for the aggregates. The equivalent radius is calculated from an aggregate's projected cross section averaged over many orientations [22].

The charge and standard deviation shown for aggregate grains follow the same trends as for spherical grains, with both the charge mean and variance proportional to the equivalent radius. For illustrative purposes, we present the results from PPD plasma conditions. The average charge and standard deviation, calculated over 90,000 time steps using PPD plasma conditions, are shown in Figure 5a and 5b for spheres (filled circles) and aggregates (open circles). However, the average charge on the aggregates tends to be greater than that for spheres (Eq. 5) since the aggregates have a greater surface area. As shown in Fig. 5c, where the aggregate charge is normalized by the charge on a sphere with $R = R_\sigma$ the difference is greatest for aggregates comprised of smaller monomers, as these aggregates tend to have the greatest surface area. The increased charge for aggregates causes the standard deviation, as a fraction of the total charge, to be smaller than that predicted using the equation for spherical grains, (Fig. 5b). However, comparing the standard deviation to that for an equivalent sphere shows that there is wide variation in σ, with the deviation tending to increase with the aggregate size (Fig. 5d).

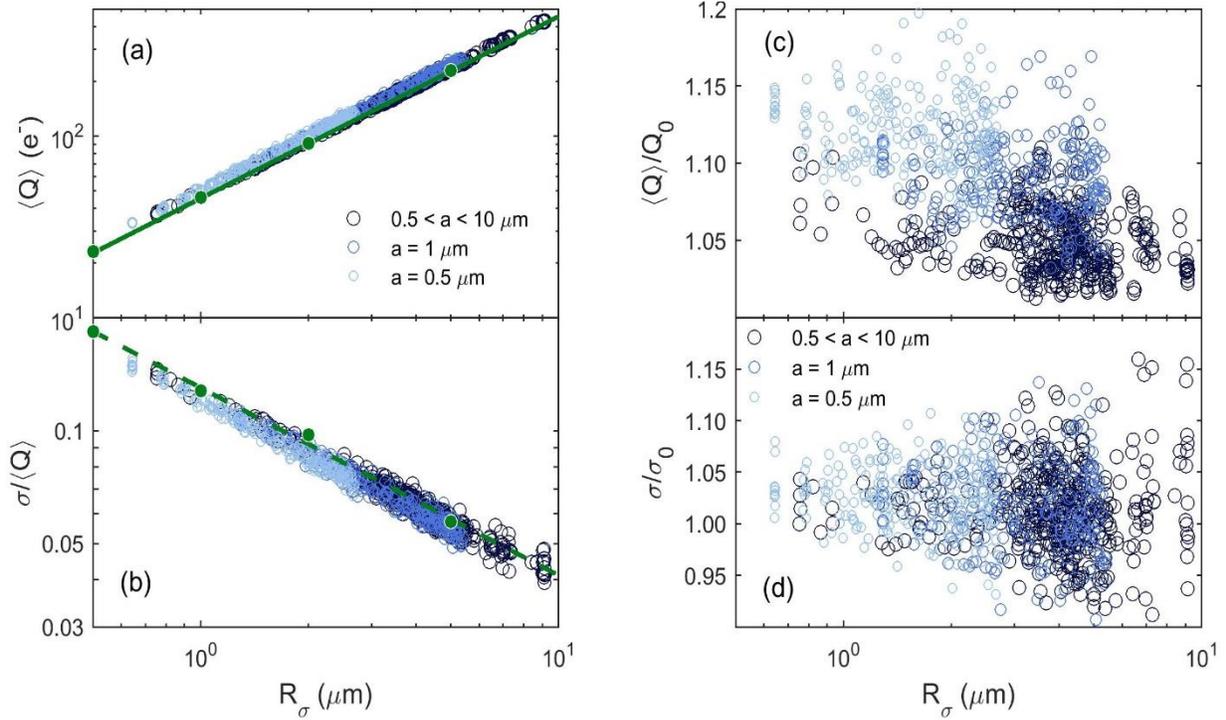

Figure 5. Average charge (a), and standard deviation of charge as a fraction of the average charge (b) for aggregates of spherical monomers with radius $a$. These values are normalized by the values predicted for spheres in (c) and (d). Data calculated for aggregates using the DSM are indicated by the open circles, while the data calculated for spheres are shown by filled circles. The lines in (a) and (b) are theoretical predictions for spherical grains. All data shown above were generated for PPD plasma conditions.

The effect of the increased charge can be seen in the fluctuation times measured for the aggregates. The growth and dissipation curves for aggregate grains in the two different plasma conditions are shown in Figure 6, along with the curves predicted for an equivalent spherical grain with radius equal to $R_\sigma$. In both cases, the growth and dissipation curves lie well below the predicted values. The fluctuation times calculated from the intersection of the growth and dissipation curves for aggregates, $\tau_c$, are shown in Figure 7 for both LAB and PPD plasma conditions, along with $\tau_f$ calculated for an equivalent sphere using Eq. 12. It is interesting to note that although the charge on aggregate grains may be increased by as much as 20% compared to the charge on spherical grains, the calculated fluctuation times tend to be shorter by a factor of ~3.

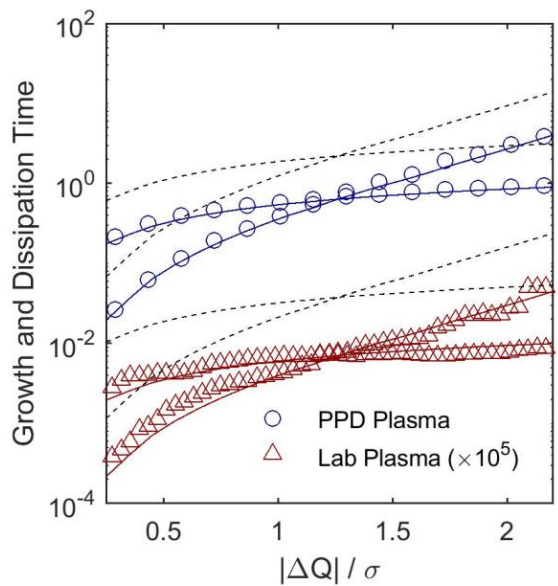

Figure 6. Characteristic times for growth and dissipation times of fluctuations on aggregate grains of 10 monomers with radius $a$ = 500 nm in two different plasma environments. The dashed lines provide analytic results for an equivalent sphere, while the solid lines are analytic fits using $\tau_c$ determined from the point where the growth and dissipation curves intersect.

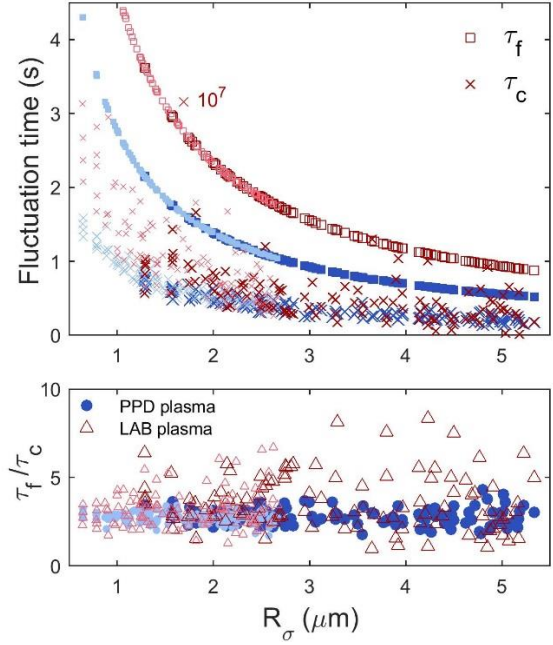

Figure 7. (Color online) Fluctuation timescales for aggregates calculated from Eq. 12, $\tau_f$, and from the intersection of the growth and dissipation curves, $\tau_c$, for (red, open symbols) LAB plasma conditions and (blue, filled symbols) PPD plasma conditions. The larger, darker symbols are for aggregates with $a = 1$ µm, and the smaller, lighter symbols are for aggregates with $a = 500$ nm. The fluctuation times for the LAB plasma are multiplied by $10^7$.

## V. EFFECTS OF DSC FLUCTUATIONS ON DYNAMICS AND AGGREGATION

The primary difference between our method and previous methods treating discrete fluctuations of charge on a spherical grain is the asymmetry in the electrostatic potential. The deviation from spherical shells of constant potential can be quite significant, resulting in a strong non-uniformity on the charge distribution on the surface of the sphere. Figure 8a shows the contour lines for a 100 nm grain with five electrons on its surface, charged in PPD plasma conditions yielding an average grain charge of 4.7 e⁻. While the resolution used is $n = 10$ patches, only five of the patches carry a charge. A similar plot is shown for the same 100-nm grain under LAB plasma conditions where the average grain charge is 214 e⁻ (Fig. 8b). In this case the charge on each patch ranges from 15 e⁻ to 30 e⁻, and the deviation from circular contour lines is much less pronounced.

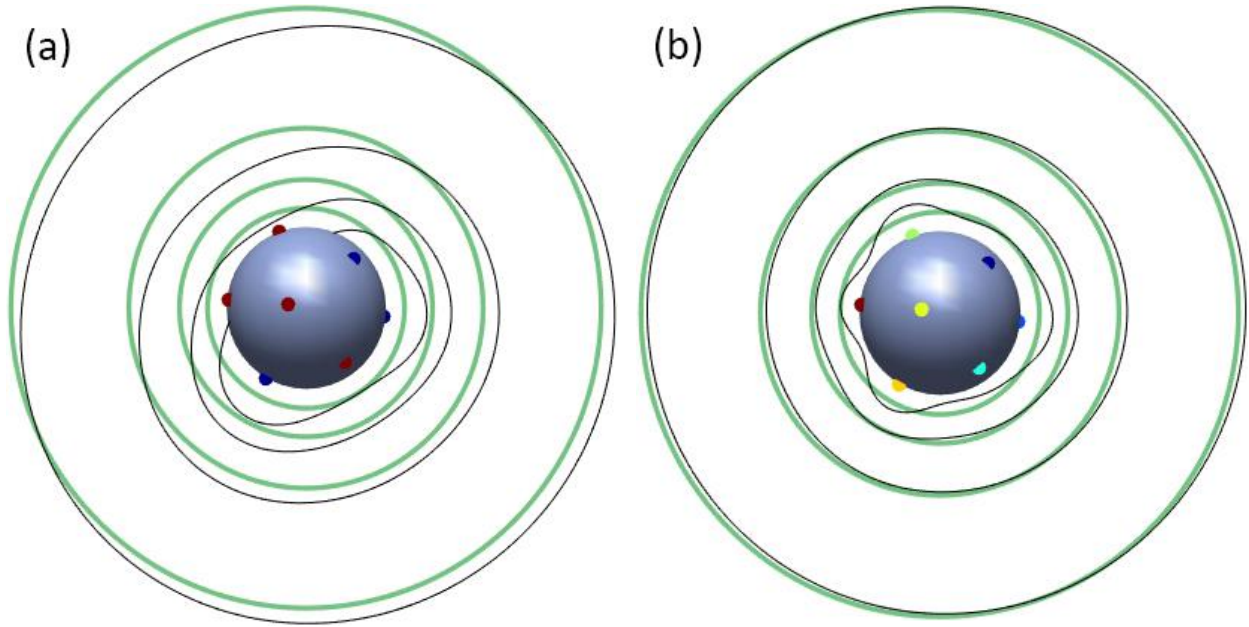

Figure 8. Contour lines for the electrostatic potential. Green lines show the potential contours for a point charge centered at the origin, black lines show the potential contours calculated for charge located at each patch point. a) Grain charged in PPD plasma condition where the average grain charge is 4.7 e$^-$. Here five of the patches, designated by the blue dots, have one electron. The patches indicated by the red dots have no charge. b) Grain charged in LAB plasma conditions with an average charge of 214 e$^-$. The patch charges range from 15 e$^-$ (red) to 30 e$^-$ (dark blue).

Previous studies have shown that the distribution of charge over an aggregate surface changes the morphology of grains produced through collisions, primarily due to rotations of the aggregates caused by the electrostatic torques [23], [24]. Thus it is of interest to determine the effects of the fluctuating charge distribution on aggregate growth. The effect on the dynamics is determined by the relative charging timescale and the particle interaction time during collisions. If the charge fluctuation time is long compared to the interaction time, then the charge can be considered to be constant during the interaction. This was the case modeled in [12] for a PPD plasma using a model for continuous stochastic charging. However, this study did not take into account the charge distribution on spherical monomers, which were treated as point charges. If the charge fluctuation time is shorter than or comparable to the interaction time, then the charge will continuously change as the two particles approach, altering the forces and torques acting on the two particles. This is the case for the LAB plasma conditions.

Here we examine aggregation in the two different plasma environments, modeling the growth through the addition of single spherical monomers. A target particle is placed with its center of mass at the origin, and new particles are projected from a random direction towards the origin plus an offset. The average charge on the grains is estimated for the given plasma conditions by balancing the electron and ion currents to determine the grain potential. Each particle is then charged employing the DSC algorithm for a sufficiently long time to cover several charge fluctuations of magnitude 1$\sigma$. Since we want to detect events which lead to collisions, the incoming particle is given a velocity just large enough to overcome the Coulomb potential barrier at a distance $d = R_{\sigma 1} + R_{\sigma 2}$ calculated using the *average* charge of each particle.

In the PPD environment we model the dynamics and growth of silicate grains with mass density $\rho$ = 2500 kg/m$^3$ using monodisperse spheres with $a$ = 100 nm. In this case, $\tau_c$ is on the order of seconds while the interaction time is on the order of milliseconds. The charges on the particles are held fixed during the interaction, but a random deviation from the average charge is obtained by running the charging code for 10,000 time steps. In the LAB plasma, we consider 25 nm melamine formaldehyde spheres with mass density $\rho$ = 1500 kg/m$^3$. The fluctuation time is on the order of 100 ns which is comparable to or shorter than the modeled interaction time. Thus the charge is allowed to change during the interaction by calling the DSC algorithm and allowing fluctuations to occur (adding one electron or one ion each random time step) until the elapsed charging time equals the dynamic time step.

Three different models of charge interactions are examined and then compared to the coagulation of uncharged particles, as illustrated in Figure 9:

> Neutral: The aggregates are not charged.
> Average: The time-averaged charge on each aggregate is used – no stochastic fluctuations are considered.
> Sphere: The charge fluctuates stochastically, but only the average charge on each monomer is considered (i.e. spherical monomers are treated as point charges)
> Patch: The charge fluctuates stochastically, and the surface charge on each patch is considered in calculating the forces and torques (therefore spheres will have higher-order multipole moments).

To speed up the calculations, outside of a distance of $4R_1$, where $R_1$ is the maximum radial extent of the target particle, multipole expansions of the particles' potentials (up to the quadrupole terms) are used to calculate the electrostatic forces and torques acting on each aggregate. Inside this distance, either the charge on each monomer (*Average*, *Sphere*) or the charge on each patch (*Patch*) is used to calculate the electric fields [24].

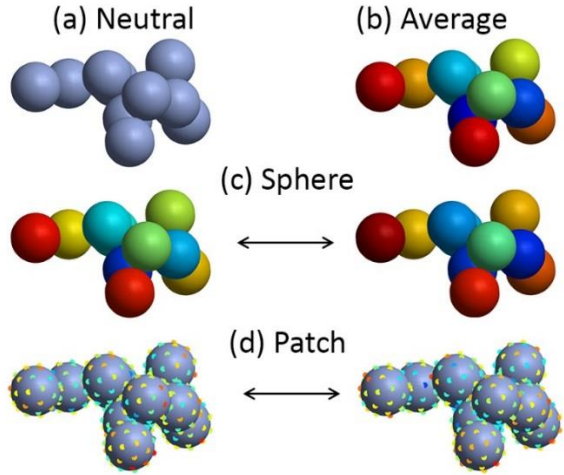

Figure 9. (Color online) Four different charging cases considered in the aggregate dynamics. (a) All particles are uncharged, (b) the time-averaged charge is applied to each spherical monomer, (c) the total charge on each monomer fluctuates in time, (d) the charge on each patch fluctuates in time. For the spheres, red indicates a large (negative) charge and blue indicates a small (negative) charge. For the patches, red indicates the largest negative charge, and blue is the largest positive charge.

After each collision, the charge on the new aggregate is obtained by running the DSC algorithm for a time long enough to cover several fluctuations of magnitude 1σ. A new monomer (with randomized charge) is set at a random incoming direction and shot towards the target. Aggregates are built up to 25 monomers in size, with more than 100 aggregates built for each case.

Figures 10-12 compare the physical characteristics of the aggregates formed using the different charging cases in both plasma environments. In all three figures, the data points represent the average quantity for all aggregates with $N$ monomers. One characteristic of particular interest for aggregate growth is porosity, or the amount of empty space, as this controls how well an aggregate couples to the gas and the surface area available for holding charge. The equivalent radius can be used to define the compactness factor $\Phi_\sigma$ which is the ratio of the volume of the constituent monomers to the volume of a sphere of radius $R_\sigma$ [22]. Thus a sphere has a compactness factor of one, while $\Phi_\sigma < 1$ for a fluffy, porous aggregate. As expected, charged grains are "fluffier" than uncharged grains as seen by the decreased compactness factor (Fig. 10), and the aggregates built with stochastic variations (*Patch* and *Sphere* cases) have the lowest (though indistinguishable) compactness factors. Plots of the average aggregate size as a function of the number of monomers are shown in Figure 11. Although $R_\sigma$ is almost identical for each of the four cases on the scale shown, the maximum radial extent of the aggregates, $R$, is quite different. The differences between the different charging models are more clearly seen in plots of $R_\sigma /R$ (Figure 12), which is used as a proxy for the aspect ratio of the aggregates. In both the LAB plasma and PPD plasma conditions, taking into consideration the fluctuations at each point on the surface (*Patch*) resulted in aggregates with the largest aspect ratios. It is of interest to note that in the LAB plasma, where the fluctuation

time is shorter than or comparable to the dynamics time, the difference between the time-averaged charge (*Average*) and the charge on each monomer (*Sphere*) is small.

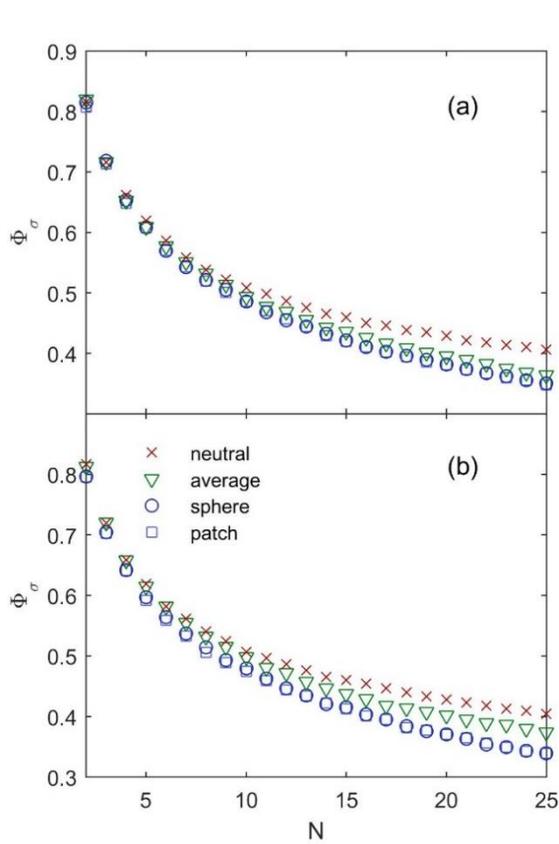

Figure 10. Compactness factor as a function of the number of monomers for (a) LAB plasma and (b) PPD plasma.

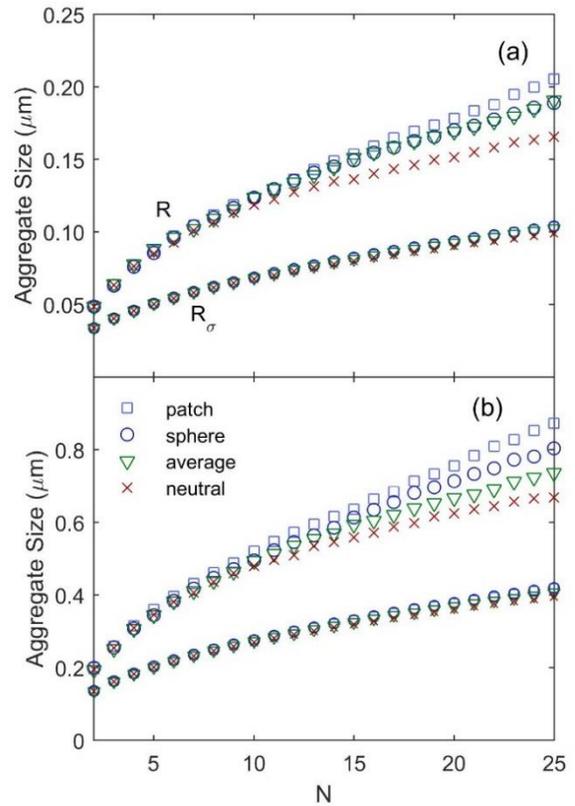

Figure 11. Aggregate size as a function of the number of monomers $N$. The upper curves designate the average maximum radial extent, $R$, measured from the center of mass, while the lower curves show the average equivalent radius $R_\sigma$. (a) LAB plasma, (b) PPD plasma.

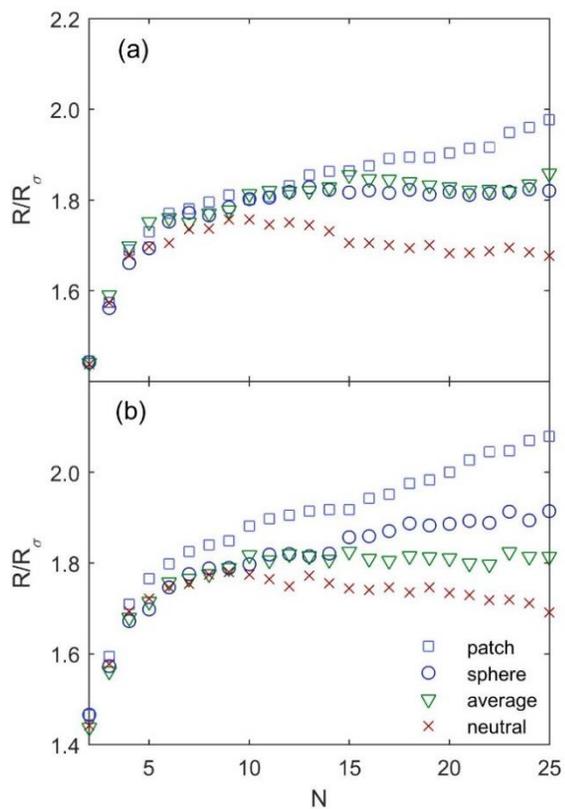

Figure 12. Ratio of the maximum radial extent to the equivalent radius as a function of the number of monomers for (a) LAB plasma, and (b) PPD plasma.

Sample aggregates from the *Patch*, *Average*, and *Neutral* cases are shown in Figure 13, which illustrates the difference in aspect ratios. Here each aggregate has been rotated so that the direction of maximum extent (as measured from the COM) lies along the x-axis. As shown, the aggregate built with the *Patch* charge variations is much more elongated (Figure 13a).

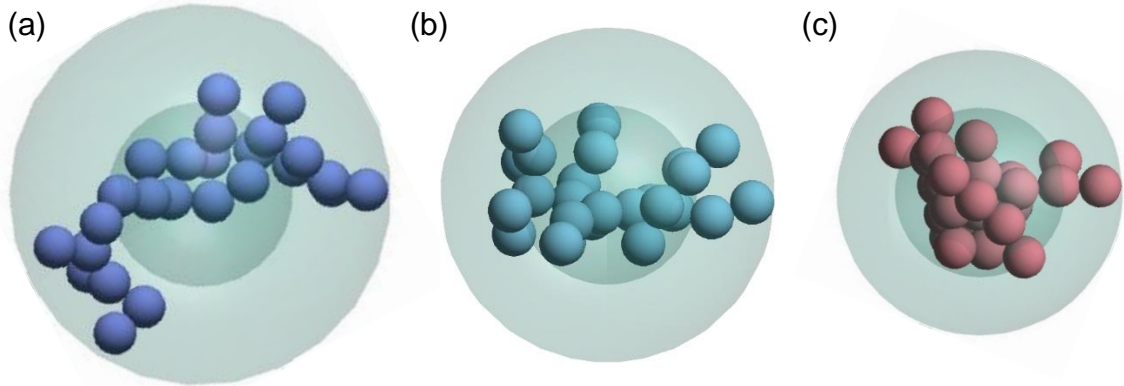

Figure 13. Comparison of aggregates built using the different charging models. All aggregates have 25 monomers. a) *Patch* charge with DSC, b) constant *Average* charge, c) *Neutral*. The inner sphere shows the extent of $R_\sigma$, which is approximately equal for the three aggregates, while the outer sphere indicates the maximum radial extent.

## VI. CONCLUSION AND DISCUSSION

We have presented a numerical model that allows examination of discrete stochastic charge fluctuations on the surface of aggregate grains and determines the effect of these fluctuations on the dynamics of grain aggregation. The DSM considers additions of single electrons or ions to patches on the surface of an aggregate based on calculated electron and ion currents to the surface points. The model recovers the results previously derived for spherical grains, where the sphere was treated as an isopotential surface [2]–[5]. It is shown that the mean and standard deviation of charge on aggregate grains follows the same trends as those predicted for spheres having an equivalent radius, although aggregates may exhibit larger variations from the predicted value for a given equivalent size (Figure 5). Although the difference in charge is on the order of ~10%, the time scale of fluctuations of aggregate grains tends to be shorter than those for spheres with the same equivalent radius by a factor of ~3, as shown in Figures 6 and 7.

While it is generally accepted that the distribution of charge over the irregular surface of an aggregate grain can greatly influence its growth process, here we show that an uneven distribution of charge on a spherical grain due to charge fluctuations (Figure 8) can also impact both the growth process and the physical characteristics of the aggregates. In particular, charge fluctuations tend to produce aggregates which are much more linear or filamentary (see Figures 12 and 13). Recent experiments have shown that ice aggregates formed from water vapor injected in RF discharges vary in both size and aspect ratio as the background gas pressure and type of gas is varied [25], [26]. Since the charging timescale depends on the ratio of electron to ion mass as well as the plasma density, stochastic variations of the charge on water ice droplets as they condense out of the vapor may play a significant role in determining overall grain morphology. The specific relationship between the plasma parameters, time scale of the fluctuations, and time scale of the particle dynamics is the subject of a current study.

## ACKNOWLEDGMENTS

Supported from NSF/DOE Grant Nos. PHY-1414523 and PHY-1414552 is gratefully acknowledged.